\documentstyle[12pt,aasms4]{article}
\received{20 March 1996}
\accepted{13 June 1996}

\slugcomment{Accepted for publication in the Astrophysical Journal}

\lefthead{Jun and Norman}
\righthead{Radial Magnetic Fields in Young Supernova Remnants}

\begin{document}

\def\etal{{\it et al.~}}
\def\eg{{\it e.g.},~}
\def\ie{{\it i.e.},~}
\title{On the Origin of Radial Magnetic Fields in Young Supernova Remnants}{}{}

\author{Byung-Il Jun \altaffilmark{1} and Michael L. Norman}
\affil{ Laboratory for Computational Astrophysics \\ National
Center for Supercomputing Applications \\ Department of Astronomy,
University of Illinois at Urbana-Champaign \\
5600 Beckman Institute, 405 North Mathews Avenue,
Urbana, IL 61801}
\altaffiltext{1}{present address: Department of Astronomy, University
of Minnesota, 
116 Church Street, S.E., Minneapolis, MN 55455, bjun@astro.spa.umn.edu}

\begin{abstract}

  We study the radio emission from young supernova remnants by means
of 3D numerical MHD simulations of
the Rayleigh-Taylor instability in the shell of the remnant.  The
computation is carried out in spherical polar coordinates ($r, \theta,
\phi$) by using a moving grid technique which allows us to finely
resolve the shell.
Three-dimensional result shows more turbulent (complex) structures in 
the mixing region than the two-dimensional result, and the instability
is found to deform the reverse shock front.  
Stokes parameters (I,Q, and U) are computed to study the radio properties of
the remnant.  The total intensity map shows two distinctive regions
(inner and outer shells). The inner shell appears to be complex
and turbulent exhibiting loop structures and plumes as a result of
the Rayleigh-Taylor instability, while the outer shell is faint and laminar
due to the shocked uniform ambient magnetic fields. 
The inner shell resembles the observed radio
structure in the main shell of young SNRs,
which is evidence that the Rayleigh-Taylor instability is
an ongoing process in young SNRs.  When only the
peculiar components of the magnetic fields generated by the instability are
considered, the polarization B-vector in the inner radio shell is
preferentially radial with about $20 \sim 50\%$ of fractional
polarization which is higher than the observed value.
The fractional polarization is lowest in the
turbulent inner shell and increases outward, which is
attributed to the geometric effect.
The polarized intensity is found to be correlated with the total intensity.
We demonstrate that the polarized intensity from the turbulent region
can dominate over the polarized intensity from the shocked uniform
fields if the amplified field is sufficiently strong.
Therefore, we conclude that the 
Rayleigh-Taylor instability can explain the dominant radial magnetic field in
the main shell of young supernova remnants.   
However, the outer faint shell shows a dominant
tangential field orientation due to the shock-compression because this
region is 
not mixed by the Rayleigh-Taylor instability, which is contrary to
observations. Therefore, another mechanism is necessary to produce the radial
components of the magnetic field at the outer shock, which we suggest 
a clumpy medium model.

\end{abstract}


\section{Introduction} 

Radio polarization studies (\cite{mil87,dic91})
of the synchrotron emission from supernova remnants (SNR) have
revealed a curious regularity: young SNRs possess a predominantly radial
magnetic field structure, whereas older remnants display circumferential
fields consistent with simple compression of the interstellar magnetic
field by the supernova shock wave. The origin
of radial B-fields in young SNRs, however, has remained a mystery. One
suggestion by Gull (1973,1975) is that the interstellar field is "combed out"
radially by dense fingers of stellar ejecta produced by the Rayleigh-Taylor
(R-T) instability acting on the decelerating contact surface between the
stellar ejecta and swept-up ambient gas. This is the third in a series of
papers in which we explore this suggestion via multidimensional numerical
MHD simulations. 

In a first paper (\cite{jns95}; Paper I) we simulated the
classical R-T instability in incompressible magnetized fluids and showed
that strong magnetic field components aligned with the gravity vector are
produced in the unstable mixing layer separating dense and light fluids. 
Initially weak magnetic fields are amplified and aligned as they are
stretched around dense, downward-plunging R-T fingers. In a second paper
(\cite{jn96}; Paper II, but see also \cite{jn95}) we simulated
the evolution of a supernova remnant in two spatial dimensions in a uniform
magnetized medium. Because the shell is strongly decelerated as it sweeps
up mass, the effective gravity vector is radially outward, and consequently
the dense R-T fingers point outward (cf. Fig. 2a). Just as in the classical
instability, our simulation showed that the swept-up magnetic field is
strongly amplified as field lines drape around these fingers, and that the
strongest magnetic field components are indeed radial. However, we could
not address in 2D whether these strong radial field components, when seen
in projection through a three-dimensional remnant, would produce a net
radial B-vector polarization, nor could we address the fractional
polarisation of the radio emission. Nonetheless, our simulation
successfully reproduced the observed clumpy radio and X-ray shell which is
due to the turbulent structure of the mixing layer.

In this paper, we remove the limitations imposed by 2D axisymmetric
calculations and present results of a fully three-dimensional numerical MHD
simulation of a supernova remnant. The simulation, described in Section 2,
employs a moving Eulerian grid technique which allows us to maintain high
numerical resolution in the region of interest--the shell. The 3D
simulation is carried out at the same numerical resolution as the 2D
simulation reported in Paper II, allowing us to directly compare the
nonlinear structures produced by the R-T instability in 2D and 3D (Section
3). In Section 4 we present our detailed study of the radio emission. The
Stokes I, Q and U parameters for the synchrotron emission are integrated
through the remnant using the self-consistently computed magnetic field
distribution and simple assumptions about the spatial distribution 
of relativistic
electrons. From these we compute the total and polarized radio intensity
and the polarization fraction of the simulated remnant which we compare
with observed remnants. Section 5 provides a further discussion on the
polarized radio emission and the origin of the observed dominant radial
magnetic field in young SNRs. Our results show that the R-T instability can
produce the radial magnetic polarization in the inner region of the remnant
if the amplified magnetic field is sufficiently strong. We then discuss the
limitations of our model and suggest the necessity for an additional
mechanism to produce the radial magnetic fields seen in the outer region of
the shell just inside the outer shock wave. Finally, our conclusions are
presented in Section 6.

\section{Numerical Simulations}

  We compute the evolution of the SNR in spherical polar geometry using
the moving grid technique discussed in Paper II to maintain high
resolution in the intershock region.
The numerical method that we use here is the
same as in our previous 2D calculation.
We take as the computational domain, $ 67.5^o \leq \theta \leq 112.5^o$
and  $ 67.5^o \leq \phi \leq 112.5^o $, and $ 0 \leq r \leq 1.05 r_{shock}$
as illustrated in Figure 1.   The intershock region
is resolved with 100x200x200 cells uniformly.
The region inside the reverse shock is
resolved with eighty ratioed radial zones (i.e.; 80x200x200 cells) 
which increase in size as $r \to 0$ in
order to avoid a too stringent Courant condition on the timestep.
Periodic boundary conditions across each angular boundary are assumed
and reflecting boundary condition at $r = 0$ is used.  The outer boundary
condition is updated every timestep with the background condition.

 The initial conditions for the physical variables in this calculation are
also identical to those used in Paper II permitting a
direct comparison of 2D and 3D results.   A supernova explosion
is initialized by depositing a kinetic energy of $10^{51}
ergs$ inside a radius of $0.1 pc$ with an assumption that the velocity is
linearly proportional to the radius.  The outer 3/7 of the ejecta mass
1.4 $M_{\odot}$ is distributed with a power law density profile $\rho
\sim r^{-7}$ and the mass distribution of the inner 4/7 of the ejecta is
assumed to be constant in order to mimic Type-Ia supernova remnants
(\cite{col69}).  We assume a uniform background density
of $1.67 \times 10^{-24} g/cm^3$ (The evolution of SNR in a nonuniform
background density has been modeled recently by Jun (1995) and will be
presented in a subsequent paper.).   
The initial temperature of the gas is assumed to be
$10^{4} K$ everywhere.  The initial magnetic field
is chosen to be tangential to the $\theta -direction$ lying in the
$r-\theta$ plane with strength $3.5 \mu G$.  
We perturb the entire space with a
random noise of 2.0\% amplitude of the density field to trigger the R-T
instability in the region near the contact discontinuity.

  The above initial state is evolved by solving the ideal compressible
MHD equations on a moving Eulerian grid using the ZEUS-3D code developed
at the Laboratory for Computational Astrophysics at the University of
Illinois, Urbana-Champaign (\cite{cla94}).  
The code uses von Neumann-Richtmyer
artificial viscosity to stabilize shocks. Fluid is advected through
the mesh using the upwind, monotonic interpolation scheme of van Leer
(second order) advection.  Magnetic fields are transported by
Constrained Transport (\cite{eva88}) modified with the Method
of Characteristics (\cite{sto92,haw95}).
Because we simulate strong shocks, the Courant time for artificial 
viscosity is the smallest time step.  Therefore we sub-cycle the viscous
routine which saves CPU by about a factor of two by avoiding redundant
calculation.  The code runs efficiently with an average zone-cycles per
second rate of $1.27 \times 10^{5}$ and about 420 MFLOPS (mega floating point
operations per second) on one processor of a Cray C90 machine.  The code
is further optimized to run in parallel using 16 processors of the Cray
C90 machine by microtasking.

\section{Growth of R-T instability in Three Dimensions} 

  Rather than repeating the basic features of the nonlinear R-T
instability in two dimensions, we focus on new results by
comparing the 3D results to the 2D simulation done with the same mesh 
resolution.   Readers are referred to Paper II
for the detailed study of the R-T instability in two dimensions.
Figure 2 shows grey scale images of density at
t=100,200,300,400, and 500 years from top to bottom.   Proceeding from
the outside inward on any panel, we see the basic structure of a SNR
shell in the decelerating phase : undisturbed ambient gas (light
grey); forward shock; shocked ambient gas (medium grey); mixing
layer/stellar ejecta (black); reverse shock and unshocked stellar
ejecta (light grey).    
The most obvious
difference is that the 3D result shows more small-scale structure in
the dense fingers of stellar ejecta.
This result was also found in our 2D and 3D simulations of 
the classical R-T instability discussed in Paper I.  In general,
the largest wavelength present at any given time is similar in both 2D
and 3D.    However, just as we found before,
the largest wavelength increases and the R-T
fingers become longer as the remnant evolves beyond the self-similar 
stage.  Another noticeable result is that some bubbles of shocked
ambient gas grow up to the reverse shock front and deform it.   
In the 3D simulation, the deformation of the reverse
shock is more severe at 100 and 200 years.  The stronger
deformation of the reverse shock at this early stage is due to the stronger
deceleration of the ejecta.  
However, the deformation of the reverse shock by the R-T bubbles
is less pronounced in a 2D simulation of higher resolution 300x400
(Paper II).
As we learned from our study of the classical R-T
instability, the kinetic energy
increases as the resolution decreases.  At low resolution, small
scales which are responsible for the dissipation of turbulent kinetic
energy are not resolved.  Therefore, we believe the deformation of
the reverse shock is likely due to a resolution effect.  
However, whether this deformation will disappear in high 
resolution 3D simulations is not yet clear.  The observational
consequence of this feature is possibly a broader radio shell.

  Figure 3 compares the angle-averaged radial profile of magnetic
field strength through the remnant in 1D, 2D, and 3D simulations.
The thickness of the mixing layer is comparable in 2D and 3D
contrary to our results of the classical R-T instability in which the 
3D simulation produced a 40 \% thicker mixing layer than in 2D.  
In the latter case, this effect is due to longer fingers being
produced in 3D than in 2D.  As 
explained in Paper I, the interaction and merging between the
fingers is reduced in 3D due to the extra degree of
freedom than in 2D.   The tendency of merging in 2D occurs due to an
inverse cascade of turbulent energy (Kraichnan \& Montgomery 1979).
Therefore, fingers in 3D can grow more easily without much loss of
kinetic energy.   Why then is this effect not seen in the SNR simulations?
The comparable thickness of the mixing layer between 2D and 3D in the
SNR simulation may be explained by three effects.
First, the wavelength of R-T fingers
under the conditions present in SNRs has grown to its maximum size.
The dominant wavelength ceases to
grow any further after about 200 years (see Fig.2).  
The maximum size of the wavelength is likely
restricted by the finite thickness of the intershock region and the finite
region between the reverse shock and the contact discontinuity.
Chevalier (1982) found that the thickness between the contact
discontinuity and the reverse shock decreases as n (the power law index
of the expanding ejected material) increases.   Later, the dominant
wavelength was found to decrease as n increases (\cite{che92}).
Second, the thickness of the heavy fluid region was assumed to be
infinite in our simulations of the classical R-T
instability, whereas it is finite in the SNR case. 
Therefore, in the former case the dominant wavelength can become larger and
larger without limit with a continuous interaction between fingers 
as it evolves.  In the SNR case, the
interaction between fingers is very weak after the R-T fingers 
reach the maximum wavelength.  As a result, the effect of 
the higher degree of freedom in 3D is not significant from the standpoint
of the mixing layer.
Third, it is likely that
the growth of the R-T finger is hindered by the high density region
behind the forward shock in the case of SNR (cf. Fig. 2 in Paper II).

  Both 2D and 3D simulations
produce a strong magnetic field layer near the contact discontinuity.
The {\it average} magnetic field strength ($\sim 1.5 \times 10^{-5}
gauss$) is higher than the shocked ambient field 
but much lower than the equipartition field estimate of $10^{-4} \sim 10^{-3}
G$ (\eg see \cite{str73,hen80,and91}) 
inferred in young radio SNR.
Numerically, the magnetic field strength in the mixing layer is found
to increase with
higher grid resolution because the effect of higher resolution is to
decrease the 
numerical dissipation and enhance the turbulent amplification of the
field (e.g. see Figs 19 and 20 in Paper I).  A resolution study in 3D
is highly desirable as a future project. 
In order to confront observations of radio SNRs on a quantitative
level, the accurate estimation of the field strength is required.

 To study the evolution of both fluid and magnetic fields in the
mixing layer, the time history of the turbulent kinetic and magnetic energy
densities is shown in Figure 4 for both 2D and 3D simulations.   The
resolution of the 2D simulation (180x200) is the same as the 3D (180x200x200).
Turbulent energy
density is defined as the peculiar kinetic energy density computed 
by excluding the
contribution from pure radial expansion of the SNR (i.e.; $E_{tur} =
{\int {1 \over 2} \rho \vert \vec v - <\vec v>_{\theta, \phi} \vert ^2 dV \over
\int dV} $) and the magnetic
energy density is also computed by taking the peculiar component
to include only the magnetic field amplified by the instabilities
($E_{mag} = {\int {1 \over 8\pi} \vert \vec B - <\vec B>_{\theta, \phi}
\vert ^2 dV \over \int dV}$) where $<\vec v>_{\theta, \phi}$ and
$<\vec B>_{\theta, \phi}$ are the angle-averaged values of $\vec v$
and $\vec B$ (the angle-averaged value is found to be close to the 1D result).
First of all, the radial components of turbulent and magnetic energy
densities are dominant in both 2D and 3D.  All components grow rapidly
at first due to rapid growth of the R-T instability during the initial
strong deceleration phase (cf. Fig.5, Paper II).  After the
first rapid rise, the turbulent energy density (particularly the radial
component) decreases dramatically due to the weak deceleration
and the shell expansion.  Therefore, the
maximum turbulent energy density is set by the highest growthrate of
the R-T instability in the early stage when the deceleration is strongest.
We find one new result here by comparing the 2D and 3D results.
The radial component of turbulent energy
density in 3D is found to be greater than in 2D until about 200 years
, and thereafter it becomes smaller.  This crossover effect
also appears in the evolution of the magnetic energy density although
crossover occurs somewhat later : at about 320 years.  
In comparison, in our simulation of the classical R-T instability
(Paper I) the energy density in 3D was found to always be greater
than in 2D except for one simulation.  This difference can be
explained as follows.   In the early stage of evolution, the interaction
and merging between R-T fingers is very strong until the
wavelength reaches the maximum size which is controlled by the
thickness of intershock region and the thickness between the contact
discontinuity and the reverse shock.  The interaction between R-T
fingers is greater in 2D due to the lower degree
of freedom (symmetry restriction) than in 3D.  Thus, the reduced
interaction between fingers in 3D results in more active growth of the
instability than in 2D.   Once the maximum size of the wavelength is
reached, this effect is no longer important.  Now more efficient
dissipation due to the small scale structures in 3D than in 2D results
in the smaller energy in 3D at late times.
About 0.57\% of total initial kinetic energy transfered into the total
turbulent energy by the end of simulation (t=500 years).   The instability
converted about 0.0011\% of the initial kinetic energy into the magnetic
energy while the total magnetic energy including shock-compressed
field reached about 0.002\% of the initial kinetic energy.  The energy
conversion into the magnetic energy by the instability seems very
low.   However, we expect from the previous studies (Paper I \& II) that
a higher resolution simulation will increase the magnetic energy 
and decrease the turbulent energy.

\section{Radio Emission} 

  In order to compute the radio emission from the remnant,
we make a few necessary assumptions.  We assume synchrotron
radiation is solely responsible for the radio emission from
young SNRs.  We take as the radio emissivity the expression (\cite{cla89})
\begin{equation}
 i(\nu) = C_1 \rho^{1 - 2\alpha}p^{2 \alpha}{(B sin \psi)}^{\alpha +
1} \nu^{-\alpha}
\end{equation}
where  $\rho$ is the gas density, $p$ is the 
gas pressure, $B$ is the magnetic field strength, $\psi$ is
the angle between the local $B$ field and the line of sight, $\nu$
is the frequency of radiation, $\alpha$ is the spectral index, and $C_1$
is a proportionality constant which depends on the normalization of
the relativistic electron energy distribution function and assumed to be 1.
Since we are not evolving the relativistic electron population
self-consistently, we assume that the spectral index is uniform
everywhere and taken as 0.6.  In the derivation of this expression, we
have made an assumption that the number density of the relativistic electrons
has a power law spectrum.  In addition, the fraction of relativistic
electrons in the fluid is an unknown constant which will determine the
amplitude of the emissivity.  A better modeling of radio emission
will require the accurate evolution of the relativistic electrons
including proper acceleration processes such as the diffusive shock
acceleration, which is beyond the scope of this paper.

\subsection{Total Intensity}

 To compute the total radio intensity (Stokes parameter I), we
integrate the emissivity along the line of sight.  Since our
numerical simulation is carried out in a 45 degree wedge in the range, 
$\theta = 67.5^o \sim 112.5^o$ and $\phi = 67.5^o \sim 112.5^o$, the 3D data
is replicated every 
45 degrees from $\phi = 0^o \sim 180^o $ along the $\phi$-direction and
then integrated along the X-direction to obtain the total
intensity (see Fig. 1; hereafter we call this the normal case.).  
Figure 5 illustrates the total radio intensity at
t=200,300,400,and 500 years.   Note that only a 45 degree sector of the
image shows the total intensity map which is truely integrated through
the entire remnant.
The radio shell imaged in the total intensity shows two distinctive regions :
the inner turbulent shell and the outer faint laminar shell.
Already at 200 years, 
the bright clumpy inner shell is clearly
seen, which is the consequence of the
amplified magnetic fields by the R-T instability.
The clumpy inner shell contains many turbulent 
structures such as loops and plumes.
These loops result from the strong magnetic fields lines
draped around the R-T fingers as they
grow.  On the other hand, the outer faint shell results from the
shock-compressed uniform ambient fields.
The maps at t= 400 and 500 years show a thicker radio shell than
at earlier times.  The loop structures are larger at later times.  
The inner turbulent shell in our simulated total intensity map is remarkably
similar in appearance to the main shell in young SNRs such as Tycho or
the ring in Cas A, which is compelling evidence for the R-T
instability in young SNRs.

 The radio emissivity depends on the angle ($\psi$) between the local
B vector and the viewing angle.  The smooth radio emission behind the
main shock is large because in this projection $\psi$ = 90.
To examine the effect of ambient magnetic field orientation on the total
radio intensity, we show the total intensity integrated along the
Z-direction (i.e., $\psi$ = 0).   
Replication of our 3D data in the $\theta$-direction
up to the pole is not a good way to produce the total intensity map
because we have no information near the pole.  Instead, we switch the
magnetic field components ($B_{\phi}$ and $B_{\theta}$) to mimic the
situation where the magnetic field is originally initialized along the
$\phi$-direction (hereafter we call this switch-B case).   
The total intensity map produced in this way is
illustrated in Figure 6b (for the amplitude, see Fig. 11).  
The total intensity near the outer shock is
very faint while the main shell remains bright.  
To assess the contribution of the instability-amplified fields to the
emission, Figure 6c shows the
total intensity map generated by the peculiar components of magnetic
fields (hereafter we call this peculiar-B case).  
The peculiar components are defined as $B_{pec} = B -
B_{avg}$ where $B_{avg}$ is the averaged magnetic field over the
angular direction.  Thus, the peculiar components are purely the 
result of the instability.
The peculiar-B case looks generally similar to the switch-B case 
but the outer shock is not seen in the peculiar-B case.   
The total intensity is
fainter overall than the switch-B case because we have removed the
contribution from the shocked mean magnetic fields.

\subsection{Polarized Intensity}

  To compute the polarized emission, we compute the Stokes Q and U
parameters as follows (\cite{cla89})
\begin{equation}
 Q(\nu) = \int i(\nu) f_o cos2\chi dl
\end{equation}
\begin{equation}
 U(\nu) = \int i(\nu) f_o sin2\chi dl
\end{equation}
where $\chi$ is the position angle of the local B field projected in the plane
of the sky, $dl$ is the increment along line of sight and $f_o$ is the
degree of linear polarization of the accumulated radiation for an
ensemble of electrons with an energy power law index $x$ that spiral
about a uniform magnetic field with an isotropic distribution of pitch
angles.  $f_o$ is given by 
\begin{equation}
 f_o = {x + 1 \over x + 7/3} = {\alpha + 1 \over \alpha + 5/3}
\end{equation}
The linearly polarized intensity and the net polarization angle
are obtained by
\begin{equation}
 P = \sqrt {Q^2 + U^2}
\end{equation}
\begin{equation}
 \chi_0 = {1 \over 2} tan^{-1}{Q \over U}
\end{equation}
respectively.

Figure 7 shows the polarized intensity (top panel) and the polarization
B-vector (bottom panel) at t = 500 years (for the amplitude, see
Fig. 11).  The vector length is
proportional to the polarized intensity; the sign of the
vector direction is arbitrary.    The polarized
intensity is found to be very high in the outer shell where the
polarization B-vector is dominantly tangential.  
The dominant tangential magnetic field is simply due to
the initial direction of the ambient magnetic field.   Because the
mixing layer does not extend all the way to  the outer shock, the
shock-compressed ambient field remains uniform.  
When a line-of-sight integration is performed, those uniform magnetic
fields dominate the polarization over the  
turbulent magnetic fields in the mixing layer.
Consequently, the polarization B-vector is sensitive to the
viewing angle.  Magnetic fields in the main turbulent shell shows some
deviation from the tangential direction, but they are not strong
enough to dominate over the shocked uniform fields (we discuss this
issue further in section 5.).
Figure 8 shows the polarization for the switch-B case.
In general, the polarization B-vector is radial and the polarized
intensity is strong in the inner main shell.   Again, the laminar post-shock
magnetic field is dominating the turbulent fields.  We then ask : what
would the remnant look like if the ambient field were completely
disordered at the outer shock by some mechanism in order to exclude
the shock-compression of uniform ambient field ?  
To approximate this without running a new simulation,
consider Figure 9, which shows the polarization
for the peculiar-B case.   We see the main shell is highly polarized; the
degree of linear polarization is about $20 \sim 50
\%$.   This range is somewhat high in comparison with the degree of
linear polarization observed in the main shell of Tycho's SNR, $7\%$.  However,
a disordered ambient field whose contribution we have ignored here 
would increase the unpolarized emission, thus decreasing
the fractional polarization.  Also, the polarization is likely
to decrease in a more turbulent flow which can be achieved by a
higher numerical resolution simulation.
The polarization B-vectors illustrate dominant radial magnetic
fields.   

   In general, the
total intensity appears to be well correlated with the polarized
intensity in the main turbulent shell (e.g. compare Fig.6(c) and
Fig.9).  Figure 10 shows the distribution of the polarized intensity with
the total intensity in the peculiar-B case.  The upper limit of the
slope is about 0.5 (the theoretical upper limit is 0.7 for a uniform
magnetic field and $\alpha = 0.6$).
The correlation between the polarized intensity and the total
intensity implies that the flow contains a characteristic scale 
carrying strong magnetic fields or the flow is not highly turbulent.
In our simulation (peculiar-B case), the stretched magnetic fields by the R-T
fingers are more responsible for a strong radio emission.
Therefore it is likely that the total intensity is
well correlated with the polarized intensity on the R-T finger scale as
long as the R-T fingers maintain their coherent structures containing
strong magnetic fields.
However, numerical
resolution is an important factor for the development of turbulence
because the higher resolution increases the numerical Reynolds number.
At a higher resolution, we anticipate the fractional polarization will
decrease due to a higher cancellation of magnetic field directions 
as the flow becomes highly turbulent.
It is unclear yet whether the R-T fingers will remain coherent and
carry dominant magnetic fields at higher resolution.  It would be
worthwhile to study the correlation between the polarized
intensity and the total intensity as a function of wavelength for both
the simulated data at various grid resolutions and the observed data.

  The radial distribution of total and polarized intensity and
fractional polarization are shown in Figure 11.  In each case, we show
the central slice from the images in Figure 9.   The normal
case and peculiar-B case show the thick shell near the outer shock
while it is hard to identify the shell in switch-B case.   This is because the
brightness originating from the shocked uniform field in the switch-B case
decreases with the radius.  The shell can be identified
better in the overall image of total intensity (see Fig.6b).
The normal case and switch-B case show the jump in fractional
polarization between the inner turbulent shell and the outer laminar shell.
Note
that the brightness in the peculiar-B case is lower than the normal case
because we removed the uniform magnetic field components.  Fractional
polarization is generally high in the entire region and it shows the
lowest degree in the inner turbulent shell region.   
From the turbulent region, the fractional polarization increases
outward. This tendency is observed in Cas A (\cite{and95}) and Tycho (\cite{dic91}).   This could be explained by
the geometric effect.  The integration along the line of sight near the
turbulent shell will sum up the magnetic field vector destructively
while this destruction will decrease with the larger radius due to
the smaller overlap of the turbulent region along the line of sight.

\subsection{Radio Luminosity}

 The time history of the radio luminosity is shown in
Figure 12.   The radio luminosity is computed by integrating the radio
emissivity over the computational space.  The integration is carried
out along lines of sight parallel to the X,Y,and Z-directions (see
Fig. 1), which we denote $L_x, L_y,$ and $L_z$, respectively.
Also plotted in Fig.12 are $L_{x,\delta B}, L_{y,\delta B}$,and
$L_{z,\delta B} $ which stand for the radio 
luminosity integrated along X, Y, and Z-directions,
with the peculiar components of magnetic fields.  Generally, all
curves show the same features in which the time derivative of 
luminosity changes from
positive to negative at about 300 years.  
The different amplitude of the curves is due to 
the different lines of sight relative to the ambient field, which
strongly influence the total luminosity.   The effect of a different
line-of-sight to the luminosity will become less important as the flow
becomes more turbulent and the field amplification increases.
Our 3D result agrees with our previous
2D simulation (Paper II) except for the earlier change of the
slope.

\section{Radial Magnetic Field in the Shell}

  Overall, our peculiar-B case more closely resembles observed radio
SNRs than do the cases which include the emission from a uniform
magnetic field.  Thus, the results of our polarization study issue
some constraints 
on the origin of the observed dominant radial field in young SNR.  
If viewing angle and shocked uniform ambient fields are the main factors
for the polarization as we found, the polarization B-vector of all 
young SNRs should not necessarily be radial. 
Various distributions of B-vector including
tangential fields should be observed.   However, as Milne(1987) found in his
polarization surveys of SNRs, the projected fields in all measured 
young SNRs are predominantly radial.   These apparently conflicting
results may be resolved by one or more of the following effects absent
from our simulation: (1) field amplification in the mixing layer may
be larger than we calculate; (2) the uniformity of the post-shock
field may be lower than we calculate because the ambient field is
non-uniform ; or (3) an additional mechanism (instability) operates at
the shock front which reduces the uniformity of the post-shock
field. We discuss these possibilities in turn.

  Let's first consider the magnetic field distribution in the inner
shell where the flow is mixed by the R-T instability.   Our simulation shows
that the polarization angle in the inner shell is dependent on the viewing
angle. This is because the emission from the amplified magnetic fields
in the mixing layer is not dominant over the emission from the shocked
uniform magnetic fields in the outer shell.
The degree of amplification of the magnetic fields in the
shell is limited by our numerical resolution.    The maximum field strength
obtained is only about 90 times higher than the
ambient magnetic field.  However, this amplification occurs very
locally and the angle-averaged magnetic field strength peaks at about
$1.5 \times 10^{-5} gauss$, only 5 times higher than the ambient field
(see Fig. 3). 
The observed magnetic field in young SNR is much stronger than our
simulated value.   A minimum energy estimate gives a magnetic field strength,
$B = 10^{-4} \sim 10^{-3} gauss$ throughout the shell of the remnant
although the minimum energy requirement (close to equipartition
assumption) is not well justified.
Without an assumption of the equipartition, Cowsik \& Sarkar (1980)
estimated the lower limit to the magnetic field in Cas A, $8 \times
10^{-5} gauss$, assuming that the bremsstrahlung by the electrons
responsible for the non-thermal radio emission does not exceed the
upper limit on the $\gamma$-ray flux.  Also, Matsui et al. (1984) used the
mean electron density inferred from an analysis of X-ray fluxes to
estimate the magnetic field in Kepler's SNR, $7 \times
10^{-5} gauss$ from the observed rotation measure.   These values are
weaker than the equipartition magnetic field but
certainly stronger than the shocked ambient field and our simulated value.
Now, we ask whether the polarized
intensity in the inner turbulent shell could become dominant over the
polarized intensity from the shocked uniform magnetic fields behind the outer
shock front as the amplified field strength increases.
Since the polarized intensity is a fraction of
total intensity, one can roughly write ignoring the distribution of
relativistic electrons
\begin{equation}
 {P_{c} \over P_{a}} \equiv {f_{c}I_{c} \over f_{a}I_{a}} \simeq
 {f_{c}B_{c}^{1+\alpha} \over f_{a}B_{a}^{1+\alpha}}
\end{equation}
where $P$, $I$, and $f$ represent the polarized intensity, the total
intensity, and the fractional polarization, and subscripts $c$ and $a$
denote quantities originating from the shock-compressed uniform
magnetic fields and the amplified magnetic fields by the R-T
instability, respectively.   
The fractional polarization for the uniform magnetic
field is 0.7 from equation(4).    The fractional polarization from the
amplified magnetic field in our simulation is about 0.2 to 0.5.
However, it is much lower in the real SNRs.
We choose $f_a = 0.05$ which is an
observed fractional polarization in the ring of Cas A (e.g. see
Anderson et al.(1995).  Then one obtains the critical value of
amplified magnetic field, $B_a \approx 5 \times B_c$, after taking
${P_{c} \over P_{a}} \approx 1$ and $\alpha = 0.6$.  This critical
strength of amplified magnetic field is weaker than the estimated
lower limit to the magnetic field in Cas A.  Therefore, the shocked
uniform magnetic field will not be able to dominate over the
amplified magnetic field in the inner shell if the amplified magnetic
field is higher than the critical value.  It should be noted that
this critical value is only a rough estimate because the number
density of relativistic electrons is assumed to be constant throughout the
remnant and the fractional polarization from the amplified magnetic
field is uncertain.   A possible high acceleration efficiency of
electrons at the shock 
will increase our estimated critical value of the magnetic field.
  
   In order to check the validity of the above discussion, we study the
radio emission with our simulated data under the assumption of higher
amplification of the field.  We compute the emissivity by giving
a multiplication factor 5 to magnetic field components stronger than
the shock-compressed field.  This is to mimic the situation that the
field amplification by the instability is five times higher than our
simulation.   Figure 13 shows the total intensity (top panel) and the
polarized intensity (bottom panel) computed in this way.  The inner
radio shell in the total intensity is much brighter compared to the
faint outer shell.  Bottom panel shows the polarized intensity
overlayed with the polarization B-vector.  The polarized intensity in
the inner shell now dominates over the outer shell so that 
the field distribution in the inner shell is no longer tangential.
Magnetic fields in the inner shell shows turbulent structure with
preferentially radial components while the outer shell still shows
tangential fields.  Therefore, we conclude that the R-T instability can
explain the dominant radial magnetic fields in
the inner shell provided the field amplification is high enough.

  Although the radial B-vector polarization in the inner shell can be
produced by the R-T instability, B-vector polarization in the
undisturbed region (outer shell) between the outer shock and the tip of the R-T
fingers remains dependent on the viewing angle if the ambient
magnetic fields are uniform.   Even if the ambient magnetic fields are
not uniform but tangled, B-vector polarization in the faint outer
shell should be preferentially tangential due to the shock-compression.
The undisturbed region is about as wide
as the region mixed by the R-T instability so that it will produce
an extended radio shell outside the inner radio shell with tangentially
polarized B-vector (e.g. see Figs. 6a, 7, and 13).   However, radial
magnetic fields are observed in the entire radio shell up to the outer shock in
young SNRs (\eg \cite{dic91} for Tycho's SNR).
Therefore, our results show that it is
necessary to have an extra mechanism which operates at
the outer shock front to produce the dominant radial components of
fields.   Or at least, the field should be randomized at the outer shock. 
Then, the radial field may become dominant inward from
the shock since the radial components of the field evolve as $B_r
\propto {1 \over r^2}$ while the tangential components evolve as $B_t
\propto {1 \over r}$ (cf. \cite{dui75,rey81}).
As suggested in Paper II, a clumpy medium
model or shock instability are possible mechanisms to randomise the
field.  Recently, a 
clumpy medium model has been simulated to produce encouraging results
on this issue (\cite{jun95,jjn96}).

\section{Conclusions} 

 We have carried out a self-consistent MHD simulation of a supernova
remnant in the decelerating phase ($ 0 \leq t \leq 500 yrs$) in three
spatial dimensions.  Our moving Eulerian grid technique allows us to
maintain a high spatial resolution in the shell and accurately model
the nonlinear growth and saturation of the Rayleigh-Taylor
instability at the interface between the stellar ejecta and swept-up
ISM.  Ambient magnetic fields are amplified in the mixing layer
produced by the instability by stretching and shearing.   The
resulting magnetic field distribution is used to compute synthetic
radio maps of total and polarized intensity as well as fractional
polarization which can be compared with observation.

  Our 3D MHD simulation produces a clumpy thick radio shell which
includes loop structures and plumes.   In our simulation, the
radio shell can be characterized by two distinctive regions according
to the total intensity : a bright, clumpy, inner shell corresponding
to the turbulent mixing layer and an outer faint shell
originating from the shocked uniform fields.   
The inner shell in the simulated total intensity map is remarkably
similar to the main radio shell in young SNRs such as Tycho and the
ring in Cas A, which suggests that the R-T instability in the shell of
young SNRs is indeed an ongoing process.
The map of polarization angle exhibits a dominant radial magnetic
field in a main turbulent shell when only the peculiar component of
the magnetic fields are included in the emissivity.  The degree of linear
polarization in the inner shell is found to be about $20 \sim 50\%$ which is
higher than the observed value in young SNR.   The fractional
polarization is the lowest in the turbulent region because of the
highest cancellation of magnetic fields. And it increases
outward, which is attributed to the geometric effect.
The polarized intensity
is well correlated with the total intensity.  It is possible that a good
correlation between the polarized intensity and the total intensity occurs
because the flow is not highly turbulent due to our limited resolution.
When we consider all magnetic field components in the intershock region,
the polarization B-vector in the inner turbulent shell 
is found to be viewing angle dependent because of an
undistorted uniform tangential magnetic field near the outer shock
front.  However, this is very likely due to the insufficient
amplification of the field in our simulation possibly by the limited
numerical resolution.  We demonstrated that the amplified magnetic field in the
turbulent region could dominate in the polarized intensity over the shocked
uniform field if the amplified magnetic field is sufficiently strong.
We conclude that the R-T instability can explain
the dominant radial magnetic fields in the
inner shell of young SNRs.
On the other hand, the radio polarization of outer faint shell is
found to be viewing angle
dependent if the ambient magnetic field is uniform, because that
region is not disturbed by the R-T instability.  Even if the ambient
magnetic field is not uniform but disordered on scales smaller than
the SNR's shell, this faint outer shell will always show the dominant
tangential components of  polarization B-vectors, contrary to what is
observed.  Therefore, our results suggest that an extra
mechanism is required to generate radial magnetic fields at the outer
shock front.

\acknowledgements

  We thank Roger Chevalier, John Dickel, and Tom Jones for helpful
discussions.  We are grateful for very useful
duscussions on the polarization of radio emission with Larry Rudnick.
We also thank the referee, John Blondin for useful comments.
The simulations were performed on the Cray C90 at Pittsburgh Supercomputing
Center.   B.-I. J. was partly supported at the University of Minnesota by NSF
grant AST-9318959 and by the Minnesota Supercomputer Institute.

\clearpage

\figcaption[fig1.ps]{A schematic representation of our computational domain
(bold lines) for the simulation of young SNR.  The sector $67.5^o \leq
\theta, \phi \leq 112.5^o, 0 \leq r \leq 1.05 r_{shock}$ is resolved
with 200x200x180 cells, with 200x200x100 uniform cells tracking the
shell (intershock region). \label{fig1}}

\figcaption[fig2.ps]{Grey scale images of the density field of the
intershock region. 
Each image shows the density distribution at t=100,200,300,400,and
500 years from
top to bottom. Left images are the results from a 2D simulation and
the right images are slices of 3D simulation. \label{fig2}}
 
\figcaption[fig3.ps]{Angle-averaged magnitude of magnetic field at t =
500 years.  \label{fig3}}

\figcaption[fig4.ps]{ Evolution of each component of turbulent and
magnetic energy 
density ($E_{tur} = {\int {1 \over 2} \rho \vert \vec v - <\vec
v>_{\theta, \phi} \vert ^2 dV \over \int dV} $ and $E_{mag} = {\int {1
\over 8\pi} \vert \vec B - <\vec B>_{\theta, \phi} \vert ^2 dV \over
\int dV}$).   \label{fig4}}

\figcaption[fig5.ps]{Grey scale images of total radio intensity at
four epochs : 
(a) t=200 yrs; (b) t=300 yrs; (c) t=400 yrs; (d) t=500 yrs. \label{fig5}}

\figcaption[fig6.ps]{Grey scale images of total radio intensity. (a)
normal case, (b) switch-B case, (c) peculiar-B case. \label{fig6}}

\figcaption[fig7.ps]{The polarized intensity and the polarized
magnetic field vectors for 
the normal case at t=500 years.  \label{fig7}}

\figcaption[fig8.ps]{The polarized intensity and the polarized
magnetic field vectors for 
the switch-B case at t=500 years. \label{fig8}}

\figcaption[fig9.ps]{The polarized intensity and the polarized
magnetic field vectors for 
peculiar-B case at t=500 years.  \label{fig9}}

\figcaption[fig10.ps]{Polarized intensity vs. total intensity for
peculiar-B case. 
$I_{max}$ denotes the maximum total intensity. \label{fig10}}

\figcaption[fig11.ps]{Slices of the brightness distribution of the total
intensity (left), the polarized intensity (middle), and fractional
polarization (right) at t=500 years.  Units for the total intensity and
the polarized intensity are arbitrary. \label{fig11}}

\figcaption[fig12.ps]{The time history of the radio luminosity computed
from our computational space.  $L_x, L_y,$ and $L_z$ are the radio
luminosity integrated along the X-direction, Y-direction, and
Z-direction, respectively. $L_{x,\delta B}, L_{y,\delta B}$,and
$L_{z,\delta B} $ are the radio 
luminosity integrated along X-direction, Y-direction, and Z-direction 
with the peculiar components of magnetic fields.  \label{fig12}}

\figcaption[fig13.ps]{Grey scale images of the total intensity (top)
and the polarized 
intensity overlayed with polarization B-vector (bottom) at t=500 years.
Emissivity is computed by putting multiplication factor 5 to magnetic
field components stronger than the shocked field. \label{fig13}}


\begin{thebibliography}{}

\bibitem[Anderson \etal 1991]{and91} Anderson, M., Rudnick, L.,
Leppik, P., Perley, R., \& 
Braun, R. 1991, \apj, 373, 146

\bibitem[Anderson \etal 1995]{and95} Anderson, M., Keohane, J.W., \&
Rudnick, L. 1995, \apj, 441, 300 

\bibitem[Chevalier 1982]{che82} Chevalier, R.A. 1982, \apj, 258, 790

\bibitem[Chevalier \etal 1992]{che92} Chevalier, R.A., Blondin, J.M.,
\& Emmering, R.T. 1992, \apj, 392, 118

\bibitem[Clarke \etal 1989]{cla89} Clarke, D.A., Norman, M.L., \&
Burns, J.O. 1989, \apj, 342, 700

\bibitem[Clarke \& Norman 1994]{cla94} Clarke, D.A. \& Norman, M.L. 1994, NCSA
Technical Report \#15, http://zeus.ncsa.uiuc.edu:8080/lca/zeus3d/zeus32.ps

\bibitem[Colgate \& McKee 1969]{col69} Colgate, S.A. \& McKee,
C. 1969, \apj, 157, 623

\bibitem[Cowsik \& Sarkar 1980]{cow80} Cowsik, R. \& Sarkar, S. 1980,
\mnras, 191, 855

\bibitem[Dickel \etal 1991]{dic91} Dickel, J.R., van Breugel, W.J.M.,
\& Strom, R.G. 1991, \aj, 101, 2151

\bibitem[Duin \& Strom 1975]{dui75} Duin, R.M. \& Strom, R.G. 1975, \aap, 39,33

\bibitem[Evans \& Hawley 1988]{eva88} Evans, C.R. \& Hawley, J. F. 1988,
\apj, 332, 659 

\bibitem[Gull 1973]{gul73} Gull, S.F. 1973, \mnras, 161, 47

\bibitem[Gull 1975]{gul75} Gull, S.F. 1975, \mnras, 171, 263

\bibitem[Hawley \& Stone 1995]{haw95} Hawley, J.F. \& Stone,
J.M. 1995, Comput. Phys. Comm., 89, 127 

\bibitem[Henbest 1980]{hen80} Henbest, S. N. 1980, \mnras, 190, 833

\bibitem[Jun 1995]{jun95} Jun, B.-I. 1995, Ph.D. thesis, Univ. Illinois

\bibitem[Jun \etal 1996]{jjn96} Jun, B.-I., Jones, T.W., \& Norman,
M.L. 1996, \apjl, accepted

\bibitem[Jun \& Norman 1995]{jn95} Jun, B.-I. \& Norman, M.L. 1995,
\apss, 233, 267 

\bibitem[Jun \& Norman 1996]{jn96} Jun, B.-I. \& Norman, M.L. 1996,
\apj, accepted 

\bibitem[Jun \etal 1995]{jns95} Jun, B.-I., Norman, M.L., \& Stone,
J.M. 1995, \apj, 453, 332

\bibitem[Kraichnan \& Montgomery 1979]{kra79} Kraichnan, R.H., \&
Montgomery, D. 1979, Rept. Prog. Phys., 43, 547

\bibitem[Matsui \etal 1984]{mat94} Matsui, Y., Long, K.S., Dickel,
J.R., \& Greisen, E.W. 1984, \apj, 287, 295

\bibitem[Milne 1987]{mil87} Milne, D.K. 1987, Aust. J. Phys., 40, 771

\bibitem[Reynolds \& Chevalier 1981]{rey81} Reynolds, S.P. \&
Chevalier, R.A. 1981, \apj, 245, 912 

\bibitem[Stone \& Norman 1992]{sto92} Stone, J.M. \& Norman,
M.L. 1992, \apjs, 80, 791 

\bibitem[Strom \& Duin 1973]{str73} Strom, R.G. \& Duin, R.M. 1973,
\aap, 25, 351 




\end{thebibliography}
\end{document}